\documentclass[epj]{svjour}
\usepackage{graphics}
\usepackage{amsmath,amssymb}
\usepackage{xcolor}
\begin{document}
\title{Probing nuclear forces beyond the nuclear drip line: The cases of  $^{16}$F and $^{15}$F}
\subtitle{A Tribute to  Mahir Hussein}
\author{V. Girard-Alcindor\inst{1}  
\thanks{\emph{Present address:} Technische Universität Darmstadt, Germany}
\and I. Stefan \inst{2}
\and F. de Oliveira Santos\inst{1} 
\and O. Sorlin\inst{1} 
\and D. Ackermann\inst{1} 
\and P. Adsley \inst{2} 
\and J.C. Angélique \inst{3} 
\and M. Assié \inst{2} 
\and M. Assunç\~{a}o \inst{4} 
\and D. Beaumel \inst{2}
\and E. Berthoumieux \inst{5}
\and R. Borcea  \inst{6}
\and L. C\'aceres \inst{1}
\and I. Celikovic \inst{7}
\and M. Ciemala \inst{8}
\and V. Chudoba  \inst{9}
\and G. D'Agata \inst{9}
\and F. de Grancey \inst{1}
\and G. Dumitru  \inst{6}
\and F. Flavigny \inst{3}
\and C. Fougères\inst{1} 
\and S. Franchoo \inst{2}
\and A. Georgiadou \inst{2}
\and S. Grévy  \inst{1}
\and J. Guillot  \inst{2}
\and V. Guimaraes \inst{10}
\and F. Hammache \inst{2}
\and O. Kamalou \inst{1}
\and J. Kiener \inst{2}
\and S. Koyama \inst{11}
\and L. Lalanne \inst{2}
\and V. Lapoux \inst{5}
\and I. Matea \inst{2}
\and A. Matta \inst{3} 
\and A. Meyer \inst{2} 
\and P. Morfouace\inst{1} 
\and J. Mrazek \inst{9}
\and F. Negoita \inst{6}
\and M. Niikura \inst{11} 
\and D. Pantelica  \inst{6}
\and L. Perrot  \inst{2}
\and C. Petrone  \inst{6}
\and J. Piot \inst{1}  
\and C. Portail \inst{2}
\and T. Roger \inst{1}
\and F. Rotaru \inst{6}
\and A.M. Sánchez Benítez \inst{12}
\and N. de Séréville \inst{2}
\and M. Stanoiu \inst{6}
\and C. Stodel \inst{1}
 \and K. Subotic \inst{7}
\and D. Suzuki \inst{13}
\and V. Tatischeff  \inst{2}
\and J.C. Thomas \inst{1}
\and P. Ujic \inst{1}
\and D. Verney \inst{2}
}                  
\institute{Grand Accélérateur National d'Ions Lourds (GANIL), CEA/DRF-CNRS/IN2P3, Bvd Henri Becquerel, 14076 Caen, France  
\and Université Paris-Saclay, CNRS/IN2P3, IJCLab, 91405 Orsay, France
\and LPC Caen, ENSICAEN, Normandie Université, CNRS/IN2P3 Caen, France 
\and Departamento de Física, Universidade Federal de São Paulo, CEP 09913-030, Diadema, São Paulo, Brazil
\and CEA Saclay, Irfu, DPhN, Université Paris-Saclay, 91191 Gif-sur-Yvette, France 
\and Horia Hulubei National Institute of Physics and Nuclear Engineering, P.O. Box MG6 Bucharest-Margurele, Romania  
\and  Vin\v{c}a Institute of Nuclear Sciences, University of Belgrade Belgrade, Serbia 
\and Institute of Nuclear Physics, PAS, Radzikowskiego 152, PL-31342 Krak\'ow, Poland 
\and Nuclear Physics Institute of the Czech Academy of Sciences, 250 68 Rez, Czech Republic or shortly Nuclear Physics Institute of the CAS, 250 68 Rez, Czech Republic.
\and Instituto de Física, Universidade de São Paulo, CEP 05508-090, São Paulo, SP, Brazil
\and Department of Physics, University of Tokyo, Hongo 7-3-1, Bunkyo, Tokyo 113-0033, Japan
\and Department of Integrated Sciences, Centro de Estudios Avanzados en Física, Matemáticas y Computación (CEAFMC), University of Huelva, 21071 Huelva, Spain
\and RIKEN Nishina Cente, Japan
}
\date{Received: date / Revised version: date}

\abstract{
The unbound proton-rich nuclei $^{16}$F and $^{15}$F are investigated experimentally and theoretically. Several experiments using the resonant elastic scattering method were performed at GANIL with radioactive beams to determine the properties of the low lying states of these nuclei. Strong asymmetry between $^{16}$F-$^{16}$N and $^{15}$F-$^{15}$C mirror nuclei is observed. The strength of the $nucleon-nucleon$ effective interaction involving the loosely bound proton in the $s_{1/2}$ orbit is significantly modified with respect to their mirror nuclei $^{16}$N and $^{15}$C. The reduction of the effective interaction is estimated by calculating the interaction energies with a schematic zero-range force. It is found that, after correcting for the effects due to changes in the radial distribution of the single-particle wave functions, the mirror symmetry of the $n-p$ interaction is preserved between $^{16}$F and $^{16}$N, while a difference of 63\% is measured between the $p-p$ versus $n-n$ interactions in the second excited state of $^{15}$F and $^{15}$C nuclei. Several explanations are proposed.
\PACS{
      {21.10.-k}{Properties of nuclei; nuclear energy levels}   \and
      {27.20.+n}{6 $\le$ A $\le$ 19}   \and
      {21.60.Cs}{Shell Model}   \and
      {21.10.Sf}{Coulomb energy} \and
      {21.10.Dr}{Binding energies and masses}   \and
      {25.60.-t}{Reactions induced by unstable nuclei} \and
      {25.70.Ef}{Resonances} \and
      {25.40.Cm}{Elastic proton scattering}
     } 
} 
\maketitle
\section{Introduction}
\label{intro}
The exploration of the nuclear landscape was made possible by the development of radioactive beams.  Nuclear models have been developed and tested in nuclei far away from the valley of stability \cite{hussein1991microscopic,hussein1994halos,descouvemont2013towards,hussein1996dipole}.  Approaching the shores of this landscape, the nuclear drip lines, has allowed the observation of several new phenomena, such as halo nuclei, modification of the effective nuclear interactions, rearrangement of the nuclear shells, and clustering. Beyond the drip line, unbound nuclei usually disappear in the form of waves in the sea of the continuum, i.e. large resonances. Theoretical description of the unbound nuclei, identification and understanding the role of specific parts of the nuclear forces, are still challenges to nuclear research. Recently, following Ikeda’s work on $\alpha$-clustering near $\alpha$-emission threshold \cite{ikeda1968systematic}, Oko{\l}owicz \textit{et al} \cite{okolowicz2012origin,okolowicz2013toward} proposed to generalize the Ikeda’s conjecture to all near thresholds states, including states involving unstable subsystems like di-neutron or di-proton.

\par
In this article, focus is given to the $nucleon-nucleon$ interaction, represented by the Two-Body Two-Body Matrix Elements (TBME) in the nuclear shell model. The idea is to investigate by how much the $nucleon-nucleon$ interaction changes when nuclei are fully embedded in the continuum.

\par
Experimentally, the effective $nucleon-nucleon$ interaction energy, labeled $\delta$V$^{\text{exp}}$, can be extracted from the measured binding energies (BE)  \cite{lepailleur2013spectroscopy,brenner2006valence}. For instance, the $n-p$ interaction energy is obtained from the relation
\[
\mathrm{\delta V^{\text{exp}} = BE(Z,N)+BE(Z-1,N-1)}
\]
\[
\mathrm{-BE(Z-1,N)-BE(Z,N-1)}
\]

\par
Theoretically, $\delta$V$^{\text{th}}$  has been predicted to depend on the spatial overlap of wave functions of the last particles \cite{brenner2006valence,heyde1994nuclear,ogawa1999thomas,Yuan14}. Indeed, as one approaches the drip line, radial wave functions for these particles spread and dilute further in space. As a result $\delta$V$^{\text{th}}$ is expected to decrease since the spatial overlap between a well bound and the unbound nucleon is smaller.

\par
A good way to shed light on the effect of the continuum in the $nucleon-nucleon$ interaction is to compare level schemes of mirror nuclei involving a bound and an unbound nucleus. The asymmetries observed between the mirror nuclei can be used to single out the role of $nucleon-nucleon$ interaction. In this article, we combine the results obtained from previous and new experiments performed at GANIL for the unbound nuclei $^{16}$F and $^{15}$F. In all these experiments, the resonant elastic scattering method was used with the thick target inverse kinematics technique \cite{gol1993possibility,Oli05,assie2012spectroscopy,RESFOS}.

\section{The case of $^{16}$F}
\label{16F}
The mirror pair, $^{16}$F-$^{16}$N, can be considered a perfect case to investigate the effect of the continuum in the $nucleon-nucleon$ force.  The structure of the $^{16}$N nucleus (S$_{n}$ = +2488.8(2.3) keV) is well known, and the first low lying states are well described assuming pure single-particle configurations. The structure of the mirror $^{16}$F nucleus (S$_{p}$ = -535(5) keV)  have been previously investigated and the results are presented in detail in Ref. \cite{stefan2014probing}. This experiment had the advantage of combining excellent energy resolution, high statistics, and precise energy calibration, marking a leap in quality and consistency over the precedent results obtained for this nucleus. Experimental and theoretical aspects of this study are presented succinctly hereafter.

\subsection{Experiment}
A radioactive beam of $^{15}$O ions was produced at the SPIRAL facility at GANIL through the fragmentation of a 95 AMeV $^{16}$O primary beam impinging on a thick carbon graphite production target. The ions were post-accelerated by means of the CIME cyclotron up to the energy of 1.2 AMeV. It was possible to obtain an $^{15}$O$^{6+}$ beam with an intensity of 1.0(2)x10$^6$~pps and 97(1)~\% purity. The ions were implanted onto a thick polypropylene (CH$_{2}$)$_{n}$ target. Some ions underwent proton elastic scattering and the scattered protons were detected promptly to the reaction in a 300 $\mu$m thick silicon detector that covered an angular acceptance of $\pm$1$^\circ$ downstream of the target ($\theta_{cm}=180^{\circ}$).

\par
\begin{figure}[h]
\center
\resizebox{0.45\textwidth}{!}{\includegraphics{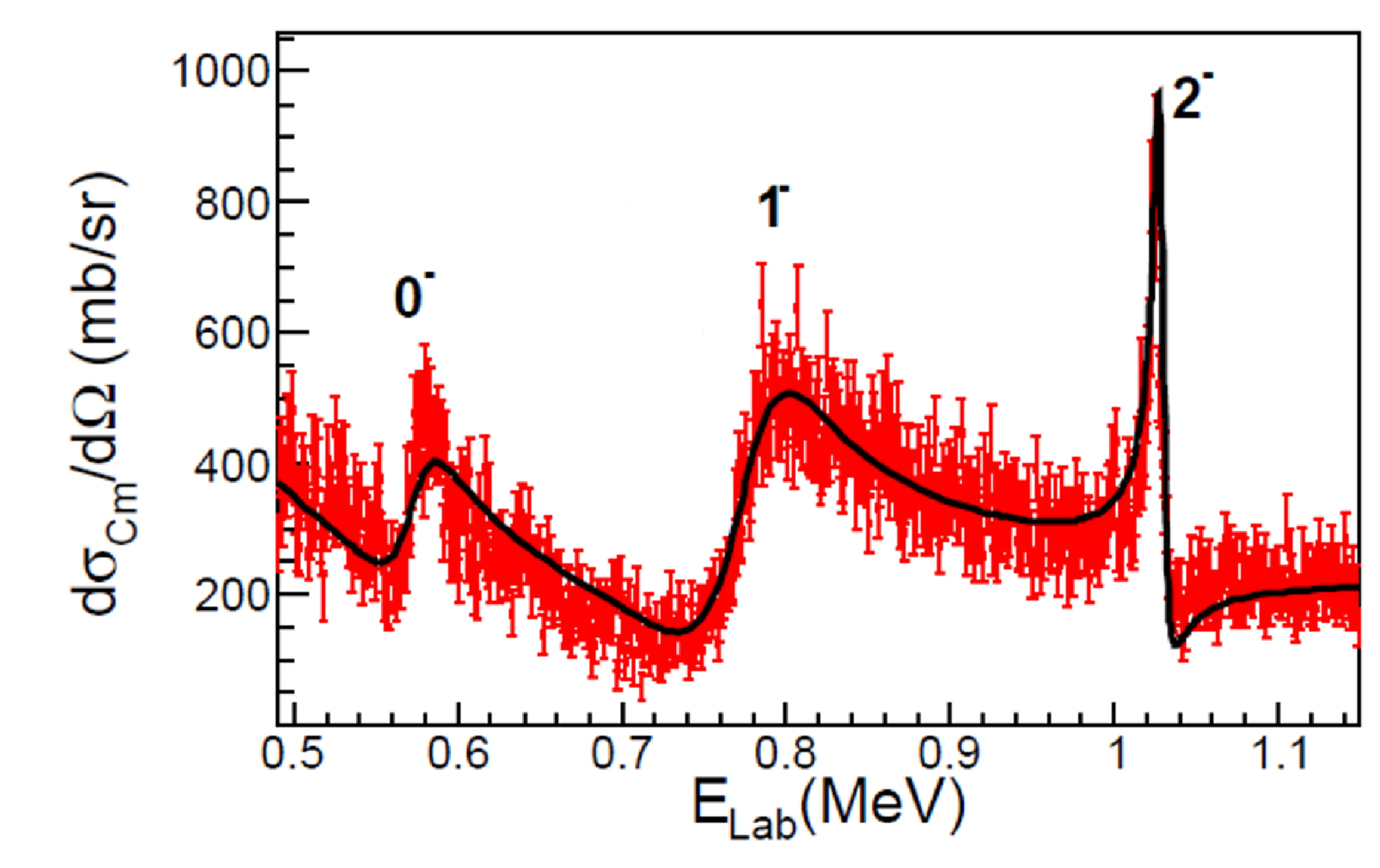}}
\caption{Excitation function of the reaction $^{15}$O($p$,$p$)$^{15}$O measured in inverse kinematics. The black line corresponds to the R-matrix fit. The presence of the 0$^-$ ground state, and two excited states $1^-$ and $2^-$ is clearly observed. The resolution of the experiment was $\simeq7$~keV FWHM in cm. Adapted from Ref.  \cite{stefan2014probing}. }
\label{16Ffig}
\end{figure}

The measured excitation function is shown in Fig. \ref{16Ffig}.  An energy resolution of  23.5(3)~keV FWHM was obtained in lab, which corresponds to $\simeq7$~keV FWHM in the center of mass. The resonances in the compound nucleus $^{16}$F ($^{15}$O +$p$) were studied through an R-matrix analysis of the spectrum.  Three resonances can be observed corresponding to the $0^-$ ground state and the $1^-$ and $2^-$ first excited states of the unbound $^{16}$F nucleus.
The measured properties are presented in Table \ref{Table1}. The deduced spectroscopic factors of the low-lying states in $^{16}$F are all close to 1. Similar values were found for the mirror nucleus $^{16}$N \cite{guo2014spectroscopic,bardayan2008spectroscopic}.
\begin{table}
\caption{\label{Table1}Measured energy, spin, width and spectroscopic factor of the low-lying states in $^{16}$F. The properties of the 3$^-$ state were adopted from Ref. \cite{lee2007low}.}
\center
\begin{tabular}{|c|c|c|c|}
\hline
E$_{x}$ (keV) &J$^{\pi}$&$\Gamma_{p}$ (keV)&S\\
\hline
0&0$^{-}$ &  25 $\pm$ 5&1.1(2)\\
193 $\pm$ 6&1$^{-}$  &  70 $\pm$ 5&0.91(8)\\
424 $\pm$ 5&2$^{-}$ &  6 $\pm$ 3&1.2(5)\\
722 $\pm$ 16&3$^{-}$ & 15.1 $\pm$ 6.7 &  $1.3(6)$   \\
\hline
\end{tabular}
\end{table}

\subsection{Effective $n-p$ interaction energies}
\begin{figure}
\center
\resizebox{0.30\textwidth}{!}{%
\includegraphics{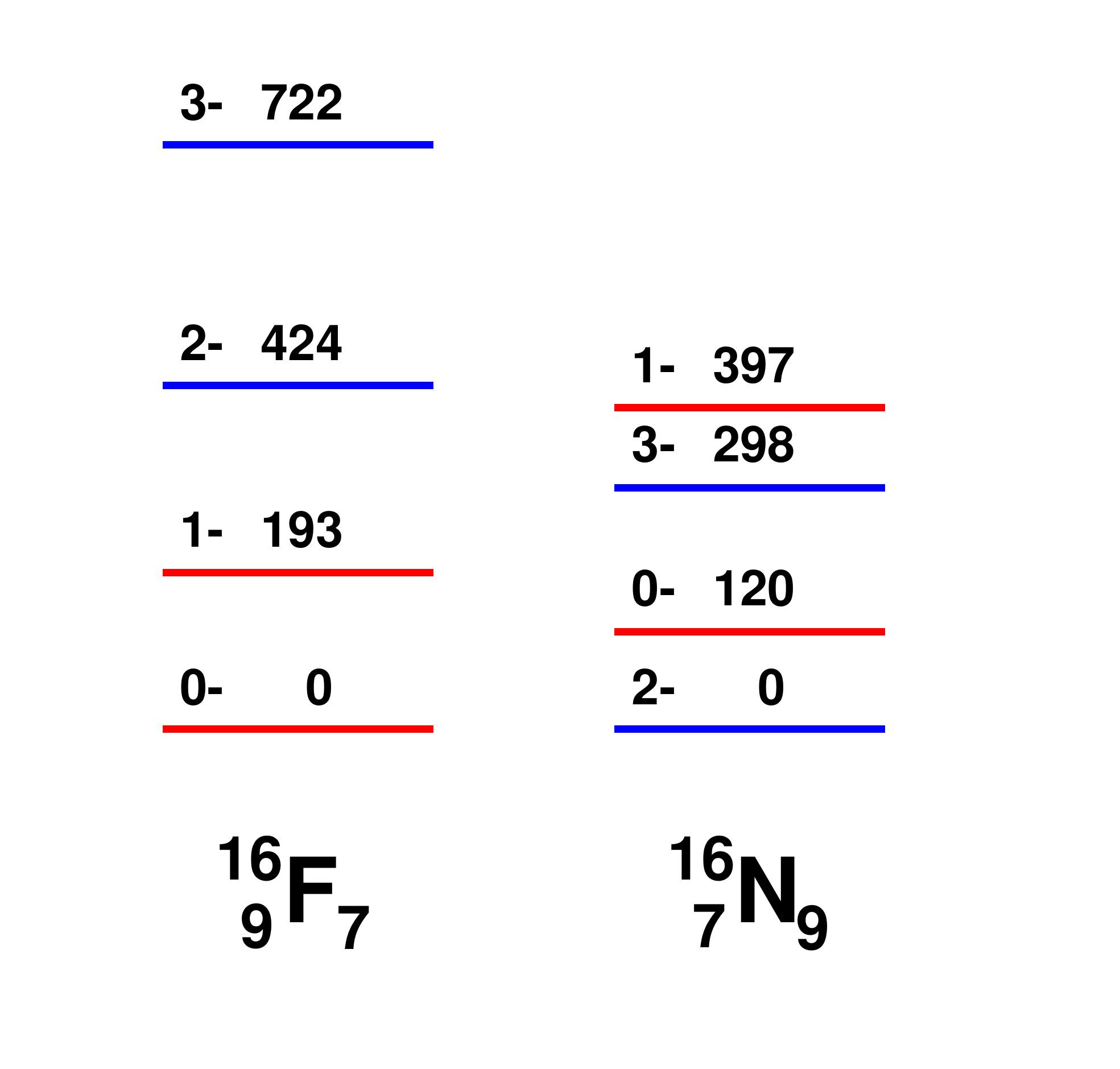}%
}
\caption{Level schemes of the two mirror nuclei $^{16}$F and $^{16}$N. The states corresponding to the 1p$_{1/2}  \otimes$ 1d$_{5/2}$ and 1p$_{1/2}  \otimes$ 2s$_{1/2}$ single-particle configurations are shown in blue and red respectively.}
\label{mirror}
\end{figure}

The level schemes of the two mirror nuclei $^{16}$N and $^{16}$F are compared in Fig. \ref{mirror}. Large differences can be observed: the ground state of $^{16}$F has $J^{\pi}=0^-$ while that of $^{16}$N has $J^{\pi}=2^-$  \cite{lee2007low}.  In $^{16}$F, compared to $^{16}$N, it looks like both 0$^-$ and 1$^-$ states are down shifted in energy by about 500~keV relatively to 2$^-$ and 3$^-$ states. 
The low-lying states of the $^{16}$N nucleus can be well described using a single-particle configuration with a closed core of $^{14}$C and a deeply bound proton in the 1p$_{1/2}$ orbital  (S$_p(^{15}$N)= +10.2~MeV) coupled either to a neutron in the  2s$_{1/2}$ orbital (S$_n(^{15}$C)= +1.22~MeV), leading to J$^\pi$=$0^-,1^-$ states, or to a 1d$_{5/2}$ neutron  (S$_n(^{15}$C*)= +0.48~MeV), forming the J$^\pi$=$2^-,3^-$ states. In the same way, (0,1)$^-$ and (2,3)$^-$ states in $^{16}$F can be described as a $^{14}$O core and a neutron in the 1p$_{1/2}$ orbital  (S$_n(^{15}$O)= +13.22~MeV) coupled either to a proton in the  2s$_{1/2}$ orbital (S$_p(^{15}$F$_{gs}$)= -1.27~MeV) or to a proton in the 1d$_{5/2}$ orbital (S$_p(^{15}$F*)= -2.79~MeV).  None of the states have the same spin, so, they don't mix, and if we neglect interaction with higher energy states of the same spin value, the spacing between the two members of a given multiplet is only due to the residual interaction. For example, in the case of the J$^\pi$=$0^-$ state in $^{16}$N ($^{14}$C+$n$+$p$), originating from the $n-p$  coupling $\pi$1p$_{1/2}  \otimes \nu$2s$_{1/2}$, the effective $n-p$ interaction is calculated with
\[
\mathrm{\delta V^{\text{exp}}(^{16}N)_{0^-} = BE(^{16}N)_{0^-} - BE(^{15}N)_{1/2^-} }
\]
\[
\mathrm{+ BE(^{14}C)_{0^+} - BE(^{15}C)_{1/2^+} }
\]
The same method was also applied to the  J$^\pi$=$1^-$ state, and to the J$^\pi$=$2^-,3^-$ states originating from the $n-p$ coupling $\pi$1p$_{1/2}  \otimes \nu$1d$_{5/2}$. 
\par
The obtained experimental $n-p$ interaction energies are given in Table \ref{Table2}.
\begin{table}
\caption{\label{Table2} Measured effective $n-p$ interaction energies. The uncertainties are lower than 10~keV on all values. Calculated from Ref. \cite{stefan2014probing,AME2016}}
\center
\begin{tabular}{|c|c|c|}
\hline
State (J$^{\pi}$)  & $\delta$V$^{\text{exp}}$ ($^{16}$N) & $\delta$V$^{\text{exp}}$ ($^{16}$F)\\
 & (MeV) & (MeV)\\
\hline
0$^-$ & -1.151  &-0.735\\
1$^-$ & -0.874  &-0.542\\
\hline
2$^-$ & -2.011  &-1.834\\
3$^-$ & -1.713  &-1.536\\
\hline
\end{tabular}
\end{table}
It is observed that the effective $n-p$ interaction changes by 40\% for the ($0^-$,$1^-$) multiplet as compared with the mirror nucleus $^{16}$N, and by 10\% for the ($2^-$,$3^-$). In all cases it is weaker in $^{16}$F, in agreement with expectations that $\delta$V will decrease as the nucleus goes in the direction of the drip line.

\subsection{Interpretation}
\begin{figure}
\center
\resizebox{0.50\textwidth}{!}{%
\includegraphics{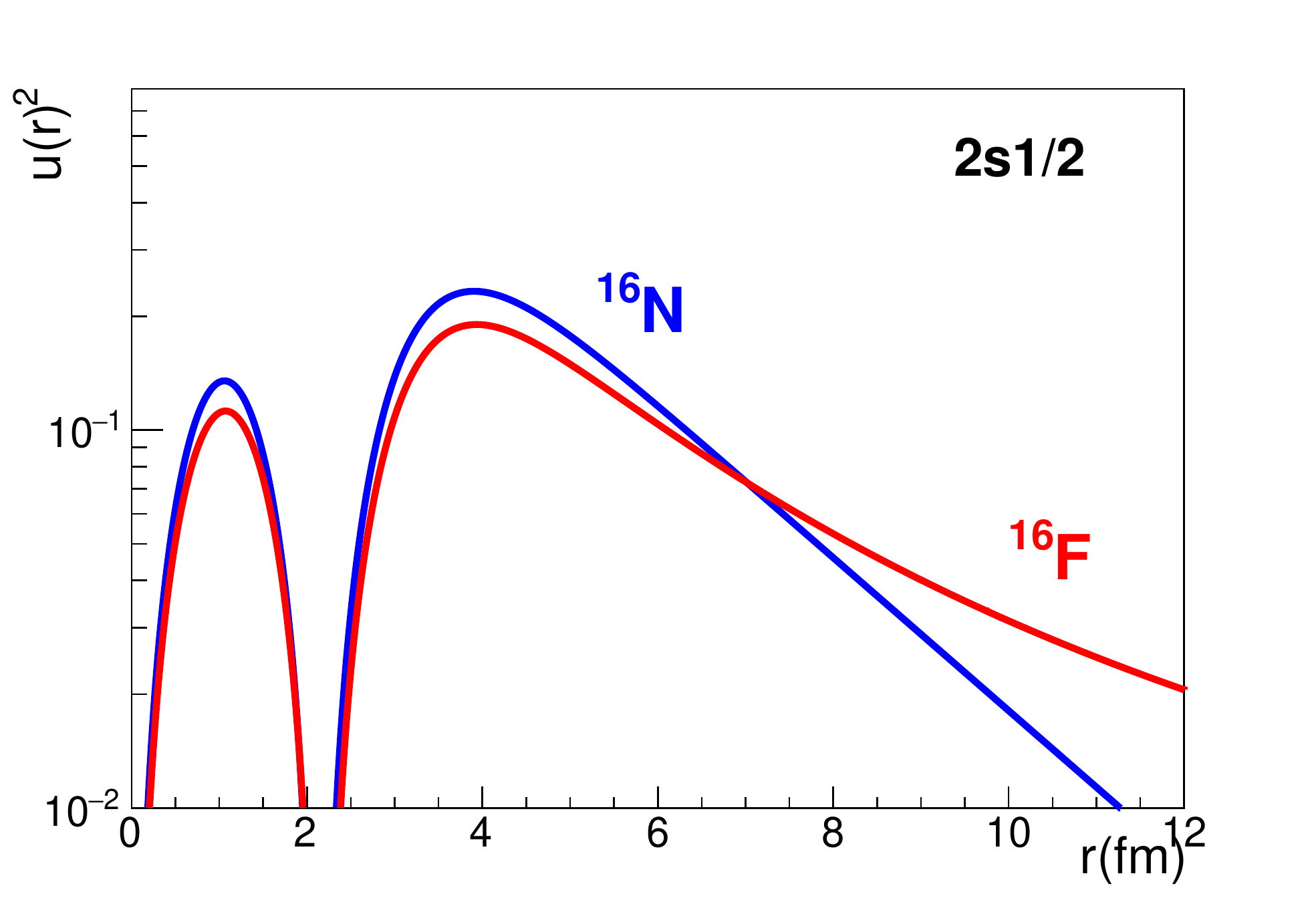}%
}
\caption{Comparison of the calculated single-particle wave functions for the $2s_{1/2}$ orbital in $^{16}$F and $^{16}$N. The difference is due to the fact that the neutron in $^{16}$N is bound whereas the proton in $^{16}$F is unbound. The wave functions have been calculated using a Woods-Saxon potential.}
\label{wf}
\end{figure}

This apparent breaking of the symmetry of the nuclear force between $^{16}$N-$^{16}$F mirror nuclei is explained by the large coupling to the continuum. The single-particle wave functions have been calculated considering a Woods-Saxon, Coulomb and spin-orbit potentials, using a standard set of parameters ($V_{WS}\approx$50-60~MeV, $a=0.6$ fm, $r_0$=1.26 fm, $V_{SO}$=6~MeV). As can be seen in Fig. \ref{wf}, the wave functions calculated for the 2s$_{1/2}$ orbit are different since the neutron is bound in $^{16}$N and the proton is unbound in $^{16}$F.
This difference modifies the overlap between the 2s$_{1/2}$  and the 1p$_{1/2}$ wave functions, and so, the effective $n-p$ interaction. This effect has been investigated in details by Ogawa \textit{et al} \cite{ogawa1999thomas} using a Woods-Saxon plus M3Y force, and by Yuan \textit{et al} \cite{Yuan14} using a monopole-based-universal interaction (VMU) in the Woods-Saxon basis. Here, we use a schematic zero-range interaction \emph{v}$~=~-a~\delta(r_p-r_n)$ to calculate the effective $nucleon-nucleon$ interaction. In the definition, $a$ is the positive strength of the interaction given in units of MeV.fm$^{-3}$, and the minus sign serves to emphasize its attractive nature. Despite its seeming simplicity, the $\delta$ interaction reproduces fairly well many properties of nuclei \cite{heyde1994nuclear}. One gets
\[
\delta V^{\text{th}}=\mathrm{-\frac{\mathit{a}_{\mathit{J}}}{4\pi} \int_0^\infty \frac{1}{\mathit{r}^2}[\mathit{u_p(r)u_n(r)]^2dr}}
\label{heyde}
\]
where $a_J$ contains the strength of the $n-p$ nuclear interaction in the state with spin $J$, and $u_n(r)$ and $u_p(r)$ are the radial wave functions calculated with a Woods-Saxon potential. By virtue of the charge symmetry of nuclear forces, the same a$_J$ coefficients have been used to calculate the interaction energies in the mirror nuclei $^{16}$F and $^{16}$N. 

\par
The calculated Reduction Factors $\frac{\delta V^{\text{th}}(^{16}\textnormal{F})}{\delta V^{\text{th}}(^{16}\textnormal{N})}$ are presented in Table \ref{factor}. These factors are of the same order as those calculated in Ref.  \cite{Yuan14}, even if they were not calculated with the same core nucleus ($^{16}$O instead of $^{14}$O). 
\begin{table}
\caption{\label{factor} Calculated reduction factor for the effective $nucleon-nucleon$ interaction.}
\center
\begin{tabular}{|c|c|}
\hline
$\delta V^{\text{th}}$  & Reduction Factor \\
\hline
$<1p_{1/2}2s_{1/2}|$\emph{v}$^{pn}|1p_{1/2}2s_{1/2}>$ & 0.66  \\
\hline
$<1p_{1/2}1d_{5/2}|$\emph{v}$^{pn}|1p_{1/2}1d_{5/2}>$ & 0.80  \\
\hline
$<(2s_{1/2})^2|$\emph{v}$^{pp}|(2s_{1/2})^2>$ & 0.33  \\
\hline
\end{tabular}
\end{table}

\par
\begin{figure}
\center
\resizebox{0.50\textwidth}{!}{%
\includegraphics{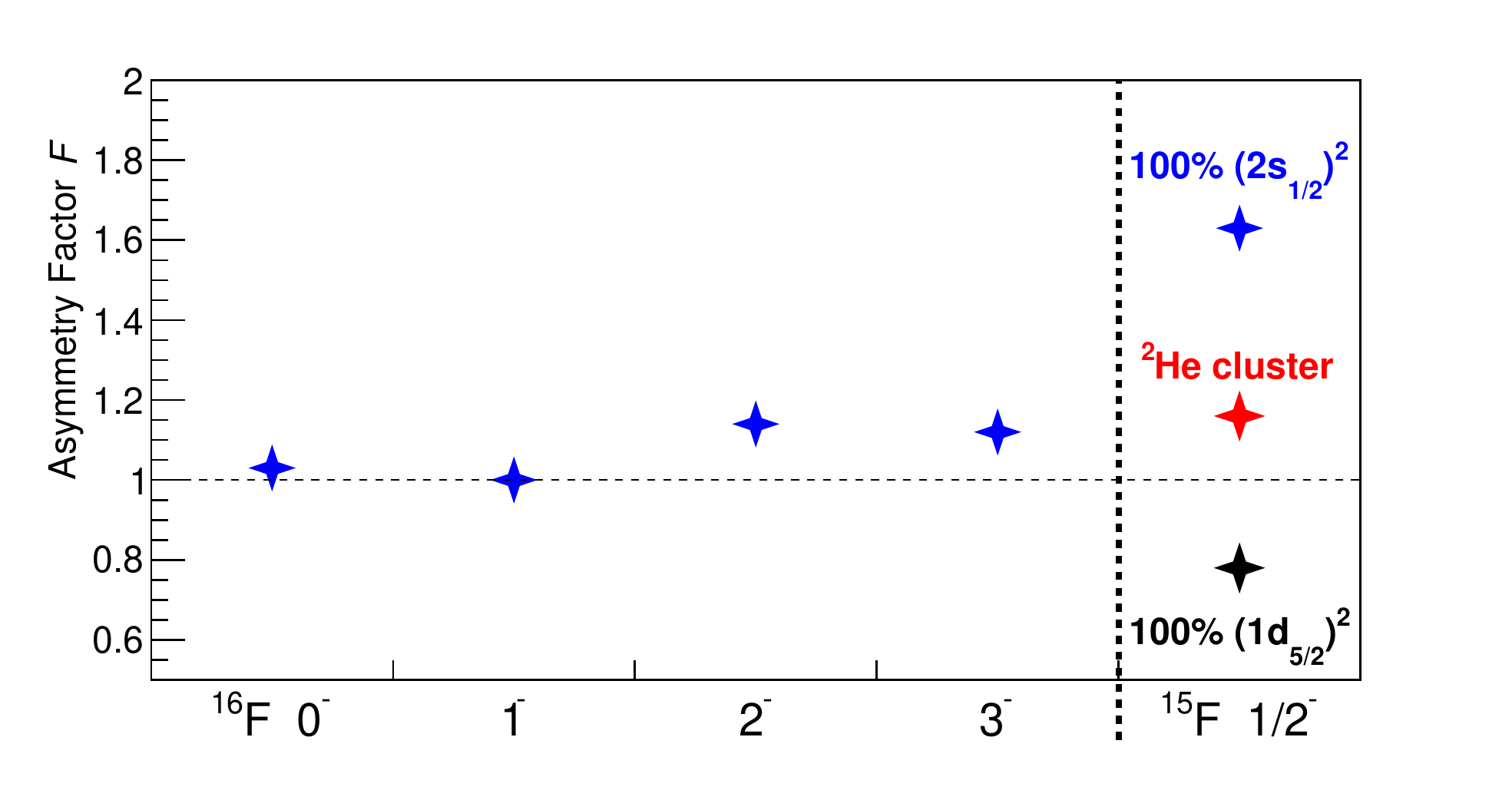}%
}
\caption{Asymmetry factors $F$ (see text) are shown for several mirror states. For the 4 states (0$^-$, 1$^-$, 2$^-$, 3$^-$) studied in $^{16}$F-$^{16}$N, the mirror symmetry is well preserved since the $F$ is close to 1.  In the case of the 1/2$^-$ narrow resonance in $^{15}$F, $F$ can be more or less close to 1.0 depending on the its structure. The case of a pure $^{13}$N+$\pi$(1d$_{5/2}$)$^2$ structure is indicated by a black star, for a pure $^{13}$N+$\pi$(2s$_{1/2}$)$^2$ by the blue star, and for $^2$He cluster+$^{13}$N core by the red star. The associated errors are much smaller than the size of the stars.
}
\label{ratios}
\end{figure}

The Asymmetry Factor $F$ = $\frac{\delta V^{\text{exp}}(^{16}\textnormal{F})}{\delta V^{\text{th}}(^{16}\textnormal{F})}  \times \frac{\delta V^{\text{th}}(^{16}\textnormal{N})}{\delta V^{\text{exp}}(^{16}\textnormal{N})}$ can be used to determine the degree of mirror asymmetry remaining after correction for wave functions differences in the mirror states. If the observed asymmetry is due solely to this effect, $F$ must be equal to 1.0. The calculated value of F for all states studied in this article are presented in Fig. \ref{ratios}. 
As can be observed in the figure, these factors are close to the expected value of 1.0 for the 4 states. Small deviations of about 10\% are observed for the 2$^-$ and 3$^-$ states, but the model used here is too simplistic to give real credit to these small deviations.  Despite the fact that the experimental $n-p$ interaction energies $\delta V^{\text{exp}}$ are reduced, the mirror symmetry is preserved. In agreement with the conclusion of Ref. \cite{ogawa1999thomas} and  \cite{Yuan14}, we confirm that the large differences in effective $n-p$ interaction energies between the mirror states are mainly explained by the different overlaps between the wave functions.

\section{The case of $^{15}$F}
\label{15F}
The isotope $^{15}$F is located two neutrons away from the proton drip line. The study of this nucleus is briefly presented below. More details about this study can be found in Ref. \cite{de2016above,degrancey:tel-00448658,valerian,girardalcindor:tel-03019149}. In addition, new results have been obtained recently and are presented here for the first time.

\subsection{Experiments}
\begin{figure}
\center
\resizebox{0.45\textwidth}{!}{%
\includegraphics{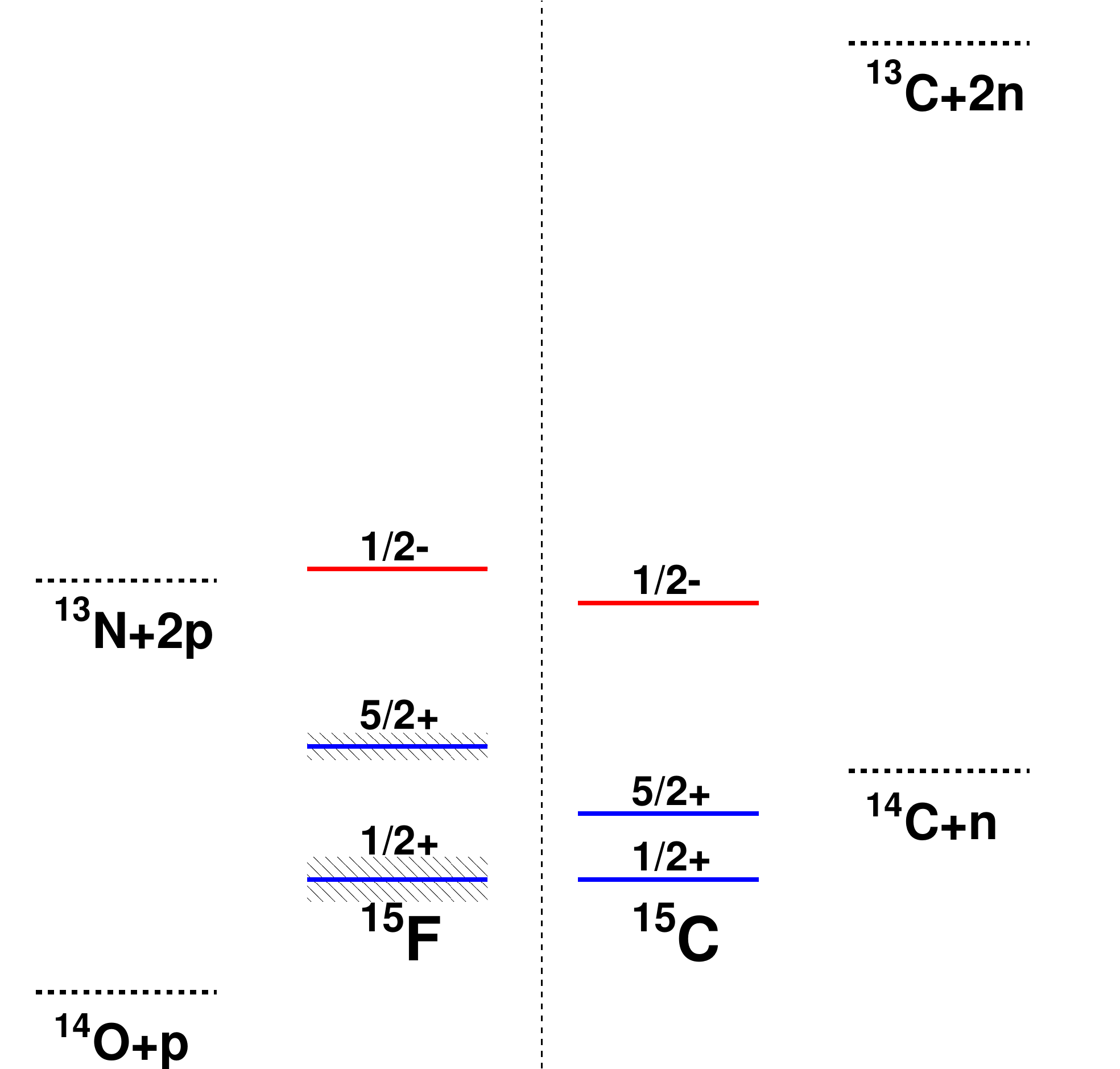}%
}
\caption{The level scheme of $^{15}$F compared to  $^{15}$C.}
\label{mirror15f}
\end{figure}

Three experiments were performed at GANIL to study the unbound $^{15}$F nucleus. The third experiment was focused on the study of the two-proton decay. The obtained results \cite{girardalcindor:tel-03019149} will be published soon in a refereed journal  \cite{valerian}. The objective of the first two experiments was the study of the structure of the low lying states. The excitation function of the elastic scattering reaction $^{14}$O(p,p)$^{14}$O was measured at 180$^{\circ}$ (c.m.), as can be seen in Fig.~\ref{15F}. It was obtained in inverse kinematics using a thick polyethylene (proton) target.
\begin{figure*}
\center
\resizebox{0.7\textwidth}{!}{%
\includegraphics{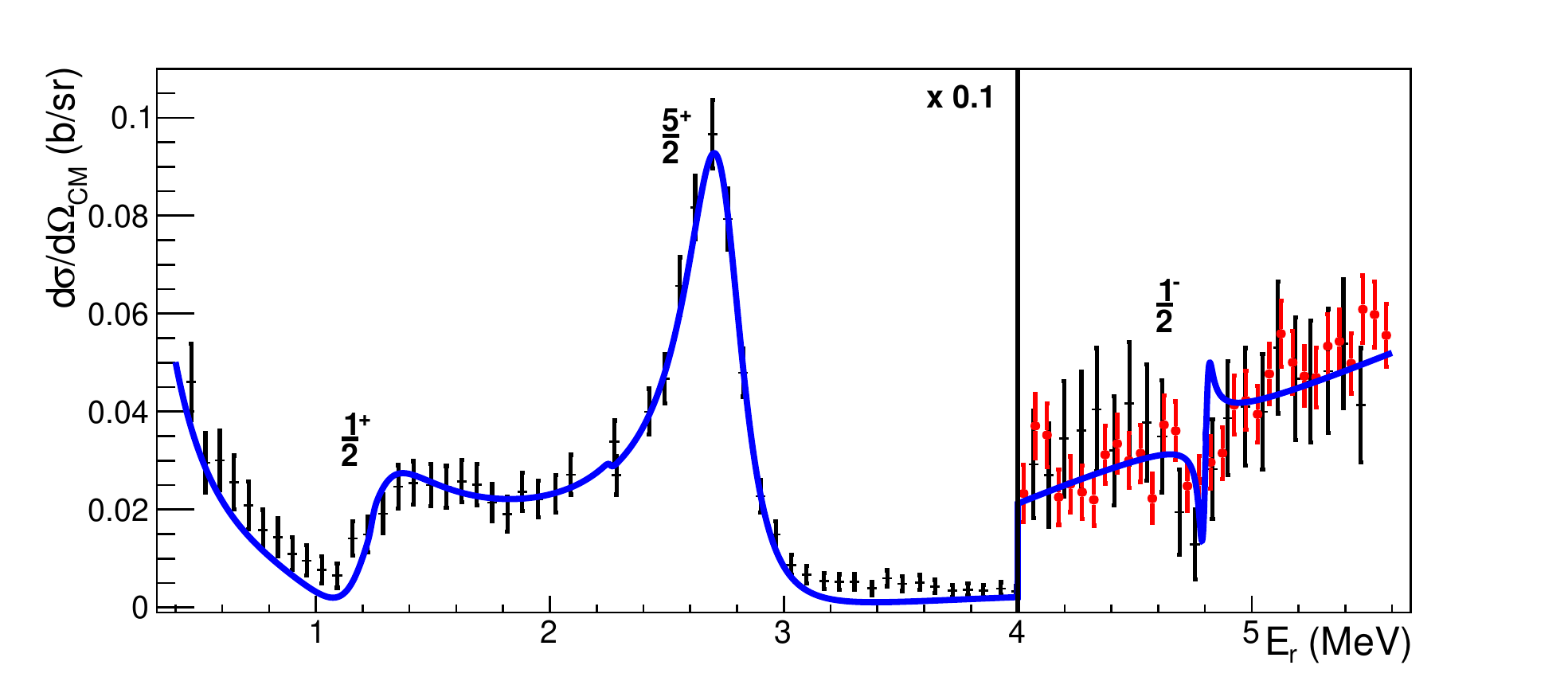}%
}
\caption{Measured excitation function of the reaction $^{14}$O(p,p)$^{14}$O. Left: Data and calculation renormalized by a factor 0.1. The experimental points in black are from Ref. \cite{de2016above}. The red points are new data confirming the existence of the dip and the $1/2^-$ resonance. The continuous blue line corresponds to the R-Matrix fit. Three unbound states are identified: The $1/2^+$ ground state, the $5/2^+$ first excited state, and the $1/2^-$ very narrow resonance. }
\label{15F}
\end{figure*}
This spectrum has high statistics, good energy resolution with 16.5(5)~keV FWHM ($\approx$~95~keV FWHM for the second experiment) and large energy covering from 0.4~MeV to 5.6~MeV (4 MeV to 5.6~MeV). The analysis of the excitation function was performed using the R-matrix method with the code AZURE2  \cite{azuma2010azure}.  The fit is shown by the continuous blue line. The measured spectroscopic properties are presented in Table \ref{Table3}, and they were used to produce the level schemes shown in Fig. \ref{mirror15f}.

\begin{table}
\caption{\label{Table3}Measured energy, spin, width and spectroscopic factors of the low-lying states in $^{15}$F. The numbers in brackets correspond to statistical and systematic uncertainties.}
\center
\begin{tabular}{|c|c|c|c|c|}
\hline
E$_{R}$ (keV) &J$^{\pi}$&$\Gamma_{p}$ (keV)&S\\
\hline
1270(10)(10)         &$1/2^+$   &  376(70)($_0^{+200}$)&0.8\\
 2763(9)(10)          &$5/2^+$    &  305(9)(10)    &$\approx$1.2\\
 \hline
4757(6)(10)           &$1/2^-$  &  36(5)(14)         &   $\approx$0.005\\
\hline
\end{tabular}
\end{table}

\par
The obtained values for the two first states are in agreement with those from the literature \cite{de2016above}. In addition, the second excited state was clearly observed as a narrow dip in the excitation function, see right side of Fig. \ref{15F}. In the mirror nucleus $^{15}$C, the second excited state has spin J$^{\pi}=1/2^-$. The shape of the dip observed in the excitation function presented in Fig. \ref{15F} could be reproduced by R-matrix calculation assuming J$^{\pi}=1/2^-$. The R-matrix fit of the excitation function was performed taking into account the experimental resolution.

\par
A second experiment was performed at GANIL in order to confirm the existence of this 1/2$^-$ state and to search for new states at higher energy. The same technique of resonant elastic scattering was used, although the measurement was made with a completely different experimental setup. A much thinner ($\approx~100~\mu$m) polyethylene target was used and the protons were detected with the MUST2 \cite{pollacco2005must2} ensemble of silicon detectors. Details and results of this experiment are presented in Ref. \cite{valerian,girardalcindor:tel-03019149}. As shown with the red dots in the right side of Fig. \ref{15F}, the new measurement perfectly confirms the presence of a dip in the excitation function. The resonance is measured at E$_R$~=~4.880(140)~MeV with $\Gamma$~=~23(10)~keV. It is in good agreement with the results of the first experiment, confirming the existence and properties of this narrow resonance.

\par
The two protons emission from this narrow resonance is energetically possible, see Fig. \ref{mirror15f}. Since there is no intermediate state accessible to $^{14}$O, it should be a direct two-proton emission to the g.s. of $^{13}$N. Considering that the available energy is only Q$_{2p}$~=~130~keV \cite{AME2016}, the branching ratio for the two proton emission is negligible ($\Gamma_{2p} < 10^{-10}$ eV), and the measured width corresponds only to the emission of a single proton towards the g.s. of $^{14}$O.

\subsection{Effective $p-p$ and $n-n$ interaction energies}
The observation of this narrow resonance in $^{15}$F is quite surprising since it is located 1.6~MeV above the Coulomb plus the $\ell$=1 centrifugal barrier for proton emission. There is no barrier to retain the proton inside the nucleus. Moreover, it is difficult to calculate the single-particle width for such a loosely bound state. The Wigner width is $\Gamma_W \approx$7.4 MeV. Experimentally, the spectroscopic factor for the one-proton emission is $S_{gs}$=0.005 (calculated with the Wigner width). It is clear that the structure of this state is far from being a proton coupled to a $^{14}$O$_{gs}$ core. It is interesting to look for the 99.5\% missing component.

\par
The situation is not completely symmetrical in the mirror $^{15}$C nucleus. The $1/2^-$ state in the mirror nucleus is at the excitation energy E$_x$=3.10~MeV, it is unbound for the emission of one neutron by 1.89~MeV, and bound for the emission of two neutrons by 6.29~MeV,  see Fig. \ref{mirror15f}. This state has also been measured as a narrow resonance, with a width of 38~keV \cite{fortune2011widths}. It is thought to be primarily an $(sd)^2$ excitation, i.e. $^{13}$C(gs)$\times(sd)^2_{0^+}$ \cite{fortune2011widths,fortune2007comment}.

\par
Theoretically, shell model calculations were performed with the code NUSHELLX \cite{brown2014shell} in the $psd$ space and with the $psdmk$ interaction \cite{millener1975particle}. It is calculated that the spectroscopic factor to the ground state is $S_{gs}=|<^{14}$O$_{gs}+p|^{15}$F$_{1/2^-}>|^2$=0.025, confirming the weak overlap with $^{14}$O$_{gs}$. 

\par
Canton \emph{et al} \cite{canton2006predicting} used the multichannel algebraic scattering (MCAS)  theory with Pauli-hindered method to predict the properties of the low-lying states in $^{15}$F.  The calculations was recently updated and several unusual narrow resonances are predicted at high energies  \cite{fraser2019mass}. A very narrow width $\Gamma$=5~keV was predicted for the $1/2^-$ state \cite{canton2006predicting}, later updated to $\Gamma$=107~keV \cite{fraser2019mass}.

\par
Fortune and Sherr \cite{fortune2007comment}, using a potential model, determined the single-particle widths for $^{15}$C. These values were scaled down to reproduce the measured widths in $^{15}$C, and the extracted dimensionless reduced widths $\theta^2$ were used to calculate widths in the mirror nucleus $^{15}$F. The results of these calculations confirmed that narrow resonances are to be expected in $^{14}$O+p, with a prediction of $\Gamma$=55~keV for the $1/2^-$ state. Refined values were later published by Fortune \cite{fortune2011widths}, with $\Gamma$=38~keV. Canton \emph{et al} \cite{canton2007canton} objected that $\theta^2$ do not necessary scale with the single-particle widths, especially when $\theta^2$ is small \cite{de1997comparison}. Moreover, the presence of a particle continuum can significantly impact spectroscopic properties of weakly bound nuclei and excited nuclear states close to, and above, the particle emission threshold \cite{MICHEL200729}, which is the case here.

\par
Gamow Shell Model in the coupled channel representation \cite{michel2009m} was also used to determine the properties of this state. The calculations were performed considering a  $^{12}$C core and three valence protons. It was confirmed that the overlap with $^{14}$O$_{g.s.}$+$p$ is very weak, with $S_{gs}$=0.0035. It was found that the $1/2^-$ state is 97$\%$ composed of two quasi-bound protons located in the $2s_{1/2}$ shell \cite{mercenne:tel-01469139}.

\par
The formalism presented in the previous section  to deduce the  $n-p$ effective interaction energies was applied to $^{15}$F(1/2$^-$) to deduce the $p-p$ effective interaction energy. We suppose that the $1/2^-$ state can be described by an inert core of  $^{13}$N$_{gs}$($1/2^-$) coupled to two protons in the $2s_{1/2}$ orbit. Shell model calculations show that the first excited state of $^{14}$O$_{1^-} $(E$_x$=5.173~MeV) is well described by a $^{13}$N$_{gs}$ plus one proton in the  $2s_{1/2}$ shell, the spectroscopic factor being S=0.70. In this case, we can write for $^{15}$F
\[
\mathrm{\delta V^{\text{exp}}_{pp} = BE(^{15}F)_{1/2^-} +BE(^{13}N)_{1/2^-}- 2 \times BE(^{14}O)_{1^-} }
\]
The $n-n$ interaction energy is determined in the same way with $^{15}$C. The obtained interaction energies are listed in Table \ref{Table15F}. However, both valence protons in $^{15}$F are charged and the Coulomb repulsion energy must be corrected for. This was calculated using the formula \cite{feynman1965feynman}
\[
\mathrm{\mathit{W} = \frac{1}{2} \int \rho(r) \phi(r) \mathit{dv}    }
\]
where $\rho(r)$ is the density of charge, $\phi(r)$ the electrical potential and $dv$ a volume element. The calculated correction is $W=179$ keV, which gives $\delta V^{\text{exp}}_{pp}$=-1.142 MeV. This value indicates that the interaction energy is weaker in $^{15}$F by a factor 1.9, once again in agreement with the expectation that the interaction decreases as the nucleus approaches, or goes further, the drip line.

\begin{table}
\caption{\label{Table15F} Measured effective interaction energies in $^{15}$F and  $^{15}$C}
\center
\begin{tabular}{|c|c|c|}
\hline
State (J$^{\pi}$)  & $\delta V_{nn}^{\text{exp}}$ & $\delta V_{pp}^{\text{exp}}$\\
 & (MeV) & (MeV)\\
\hline
1/2$^-$ & -2.127  &-0.963\\
\hline
Coulomb corrected &  -  &-1.142\\
\hline
\end{tabular}
\end{table}

\subsection{Interpretation}
If the $1/2^-$ state in $^{15}$F is described as a $^{13}$N core+ two proton in the 2s$_{1/2}$ orbital, a huge reduction factor of 0.33 is calculated, see Table \ref{factor}. This is in agreement with the results of Ref. \cite{Yuan14} where the greatest reduction factor has been observed for this TBME. Actually, their calculated value is 0.68, which is twice the value of the present work. However, it should be emphasize that their values has been calculated with a different interaction. Also, the asymmetry factor $F$, shown in Fig. \ref{ratios}, is 63\% larger than the expected value of 1. Such a large difference is significant and needs an explanation. One of them is the fact that core polarization effects, which should be included, has not been taken into account in our simple model.

\par
Also, it is possible that the $a_J$ strength coefficients are different between the two nuclei.  This may be due to the fact that the $1/2^-$ resonance in $^{15}$F is located only 130~keV above the two-proton emission threshold, whereas in the mirror nucleus $^{15}$C, the $1/2^-$ state is bound by 6.3~MeV for two-neutron emission. This asymmetry in $^{15}$F/$^{15}$C would be an indication of the generalized Ikeda's conjecture \cite{okolowicz2012origin,okolowicz2013toward}. The $p-p$ correlation might be enhanced in $^{15}$F compared to the $n-n$ correlation in the mirror nucleus $^{15}$C. To estimate the impact of this correlation enhancement, we treated $^{15}$C(1/2$^-$) as $^{13}$C + $^{2}$n where the $^{2}$n cluster is bound in a Woods Saxon potential well. We then applied the same potential to calculate the resonance energy of the mirror system $^{15}$F as $^{13}$N+$^2$He, correcting for the Coulomb energy in $^2$He ($\approx$~200 keV). The calculated asymmetry factor is $F$=1.16, as shown in Fig. \ref{ratios} by a blue star. Thus, to explain the $^{15}$F/$^{15}$C difference it would be necessary to describe the $1/2^-$ state of $^{15}$F as being 100\% composed of $^{13}$N+$^2$He cluster.

\par
Another much simpler solution would be to add to the (2s$_{1/2})^2$ wave function a part of (1d$_{5/2})^2$ component, which would have the effect of increasing the overlap of the wave functions, reducing the asymmetry. To get the asymmetry factor $F=1$, the states in $^{15}$F and $^{15}$C should be described as composed with 58\% of (1d$_{5/2})^2$ and 42\% of (2s$_{1/2})^2$. Fortune \textit{et al}  \cite{fortune1978sd} predicted 46\% and 54\% respectively, using the Lawson-Serduke-Fortune two-body matrix elements. Using these values, they predicted the energy of the $^{15}$F $1/2^-$ state at E$_R$ =4.63 MeV \cite{fortune2007comment}, which is only 127~keV below our measured energy.

\section{Conclusion}
In the present investigation we have explored the $nucleon-nucleon$ force in nuclei beyond the proton drip line. Strong asymmetry between $^{16}$F-$^{16}$N and $^{15}$F-$^{15}$C mirror nuclei was observed. For the $^{16}$F and $^{16}$N pair, the apparent breaking of the symmetry can be explained by the difference in the overlaps of the wave functions due to the coupling with the continuum. The recently measured and confirmed 1/2$^-$ state in $^{15}$F, on the other hand, shows asymmetry that cannot be explained by the difference in the overlaps of pure (2s$_{1/2})^2$ wave functions. This may be an indication of an increase in correlation between the two least bound protons due to the proximity with the two-proton emission threshold. It may also be an indication that the states are described by an almost equal mixture of (2s$_{1/2})^2$ and (1d$_{5/2})^2$ components.

\section{Acknowledgments}
We would like to thank the GANIL accelerator staff for their support. We also received support by OP RDE, MEYS Czech Republic under the project EF16-013/0001679, by LIA NuAG and COSMA, and by IRP France-Brésil.

 \bibliographystyle{epj.bst}
 \bibliography{ valerian_v4.bbl}

\end{document}